\documentclass[runningheads]{llncs}
\usepackage{amsmath,amssymb,amsfonts}
\usepackage{algorithmic}
\usepackage{graphicx}
\usepackage{textcomp}
\usepackage{subcaption}
\usepackage{xcolor}

\newcommand{\nop}[1]{}

\begin{document}

\title{On Traffic Characterization in Time-Sensitive Networking and Deterministic Networking} 
\author{Yuming Jiang}        
\institute{Norwegian University of Science and Technology (NTNU),~Norway \\
\email{yuming.jiang@ntnu.no} }
\maketitle

\begin{abstract}
The traffic of an audio/video stream in IEEE Time-Sensitive Networking (TSN) or an asynchronous traffic flow in IETF Deterministic Networking is specified by three parameters: an observation time interval, a maximum packet size, and a maximum number of packets in the interval. This traffic specification (TSpec) triggers three fundamental questions for TSN and DetNet: (1) if and how the traffic of a flow can be specified by these parameters after it traverses a system, (2) how to specify the traffic of a flow aggregate using the three parameters, and (3) how to use it in performance guarantee analysis. Among them, the literature has focused on (3), leaving the other two little investigated and largely open. This paper is intended to bridge the gap. Specifically, we introduce an equivalent traffic characterization for TSN TSpec, which is simple in  representation and simplifies analytical development. Based on this result, we establish an output characterization property for (1) and show that the output of a TSN flow traversing a bounded-delay system can be characterized by the TSN TSpec with updated parameters. For (2) we study the superposition of TSN flows and derive TSN traffic characterizations for the aggregate of multiple TSN flows. For (3), based on a relationship between the TSN traffic characterization and the well-known token-bucket traffic characterization, backlog and delay bounds are derived. Together, these results form a calculus for TSN traffic characterization. 
 
\end{abstract}

\section{Introduction} \label{sec-1}
In IEEE Time-Sensitive Networking (TSN) \cite{8021Q} and IETF Deterministic Networking (DetNet) \cite{DetNet}, the traffic of a TSN audio/video (A/V) stream or an asynchronous DetNet flow is specified by three parameters \cite{8021Q} \cite{DetNet}: 
\begin{itemize}
\item {\em an observation time interval} 
\item {\em a maximum packet size} 
\item {\em a maximum number of packets in the interval}
\end{itemize} 
This TSN traffic specification (TSpec) triggers three fundamental questions for TSN and DetNet networks: 
\begin{itemize}
\item (Q1) -- if and how the traffic of a flow can be specified by these parameters after it traverses a system;
\item (Q2) -- how to specify the traffic of a flow aggregate using the three parameters;
\item (Q3) -- how to use it in performance guarantee analysis.
\end{itemize}

Among these questions, (Q1) and (Q2) concern the closure properties of the TSN traffic model, while (Q3) concerns performance guarantee analysis more directly. As a highlight, the closure properties play a fundamental role in performance analysis because they determine whether the same traffic model remains useful after multiplexing and forwarding operations. 

Specifically, (Q1) is particularly important to understand how the characteristics of a flow alter in the network if re-shaping to the initial traffic characterization is not implemented, which can affect backlog and delay bounds at later nodes on the path. In TSN and DetNet, flow aggregation is used as an important technique to aggregate individual flows and their associated resource control to improve scaling \cite{rfc8938}. Such an aggregate is the aggregation of multiple individual flows. Since each individual flow is characterized by the TSpec, it is natural to ask how to characterize the aggregate using those parameters and apply them to the aggregate, which is (Q2). Since TSN and DetNet are intended to provide deterministic delay guarantees, (Q3) is crucial.

As is clear from the definition, the TSN TSpec implies a deterministic traffic model. In the literature several deterministic traffic models already exist and have been extensively investigated often under the {\em network calculus} umbrella \cite{Cruz91ab}\cite{Chang00}\cite{NetCal}\cite{SNC}\cite{DNC}.  Among them, the most known deterministic traffic model is perhaps the token-bucket model \cite{Cruz91ab}. The token-bucket model has two parameters, a bucket size to upper-bound the traffic burstiness and a token generation rate that upper-bounds the traffic rate. The token-bucket model has been widely adopted to study quality of service (QoS) in communication networks and systems. An example is the traffic specification (TSpec) used in the two Internet QoS architectures, Integrated Services (IntServ) \cite{BCS94} and Differentiated Services (DiffServ) \cite{Blake98}. This IntServ/DiffServ TSpec takes the form of a token bucket plus a maximum rate and a maximum packet size \cite{TSpec}.   
Extensions or generalizations of the token-bucket model and their systematic study are available in the {\em network calculus} literature \cite{Cruz91ab}\cite{Chang00}\cite{NetCal}\cite{SNC}\cite{DNC}. 
For network analysis, the output characterization property of a traffic model refers to that the same model can be used to characterize the output traffic as the input traffic, and the superposition property of a traffic model refers to that the same model can be used to characterize the aggregate of flows where each individual flow is characterized by the traffic model (Section 1.2, \cite{SNC}). They correspond to (Q1) and (Q2) respectively. For (Q3), performance guarantee analysis in terms of backlog and delay bounds has been performed under various combinations of traffic and server models in network calculus \cite{Cruz91ab}\cite{Chang00}\cite{NetCal}\cite{SNC}\cite{DNC}. 
For token-bucket traffic characterization, all these questions (Q1) -- (Q3) have been thoroughly investigated, starting from \cite{Cruz91ab}. 

However, for TSN and DetNet, while the TSN TSpec has been introduced for several years, its study is limited to (Q3). For (Q1) and (Q2), they remain open to be answered. To bridge the gap, this paper establishes, to the best of our knowledge, the first systematic calculus for TSN  traffic characterization. Specifically, we introduce a new equivalent representation of TSN TSpec based on packet arrival times, prove output characterization and superposition properties for TSN traffic characterization which are analogous to those of token-bucket traffic models, and develop backlog and delay bound analyses by establishing a formal relationship between TSN TSpec and token-bucket characterization.
Together, these results provide {\em a calculus for TSN traffic characterization} analogous to the role played by the token-bucket traffic model in deterministic performance guarantee analysis.

The remainder of the paper is organized as follows. 
In Section \ref{sec-2}, the system model and a calculus for token-bucket traffic characterization are introduced. 
In Section \ref{sec-3}, a calculus for TSN traffic characterization is developed, which includes a new representation of TSN TSpec, and output characterization, superposition property, and backlog and delay bound analysis results. In Section \ref{sec-4}, a case study is conducted to illustrate the applicability of the  developed TSN calculus. The paper is concluded in Section \ref{sec-5}. 

\section{System Model and Preliminaries} \label{sec-2}

\subsection{System Model}
We consider an initially empty TSN / DetNet system, which may be a data source, a shaper, a transmission selection algorithm, a network element, or a network of such systems in TSN / DetNet networks. By convention, a packet is considered to have arrived at (respectively, departed from) the system if and only if its last bit has arrived (respectively, departed). 

A flow in the system is a sequence of packets. Each packet $n = 1, 2, 3, \dots$ is described by a tuple $(a(n), l(n))$, where $a(n) (>0)$ represents the arrival time of the packet, and $l(n)$ its length (in bits). Packet $0$ is a virtual packet defined with $a(0)=0$ and $l(0)=0$. In TSN and DetNet, the length of each packet is not specified. Instead, a maximum packet size $S \equiv l_{max}$ and optionally a minimum packet length $l_{min}$ are included in traffic specification. Clearly there must hold $l_{max} \ge \max_n\{l(n)\}$ and $l_{min} \le \min_n\{l(n)\}$. 

The input traffic to the system is also described by a wide-sense increasing time-function, which is denoted as $A^P(t)$ if the traffic is counted using packet as the unit or as $A(t)$ if the traffic is counted using bit as the unit. Without loss of generality, we adopt that the traffic at $t$ is not counted in $A^P(t)$ or $A(t)$, i.e. in $[0, t)$. 
The traffic time-functions can be related to the sequence $\{(a(n), l(n))\}$ for the flow as follows:
\begin{eqnarray}\label{eq-connet}
A^P(t) &=& \max \{n: a(n) < t \} \\
A(t) &=& \sum_{0\le n }l(n) \mathcal{I}_{a(n) < t} 
\end{eqnarray}
where $\mathcal{I}_E$ is the indicator function, i.e., $\mathcal{I}_{a(n) < t} =1$ if $a(n) <t$, and 0,  otherwise.
Let $A^P(s,t) \equiv A^P(t)-A^P(s)$ and $A(s,t) \equiv A(t)-A(s)$, $\forall 0 \le s \le t$.

The delay $D(n)$ of packet $n (\ge 1)$ in the system is  
\begin{equation}
D(n) = d(n) -a(n).  
\end{equation}
The virtual delay $\tilde{D}(t)$ at time $t (\ge 0)$ is defined as \cite{NetCal} 
\begin{equation}
\tilde{D}(t) = \inf\{\tau \ge0: A(t) \le A^{*}(t+\tau)\}.
\end{equation}
where $A^{*}(t)$ denotes the cumulative amount of output traffic (in bits) from the system up to time $t$ (excluded). 
It has been shown that for a FIFO system, an upper bound on virtual delay is also an upper bound on packet delay \cite{NetCal} \cite{Jiang26}. 

When multiple flows are considered, a subscript is added to distinguish them. Specifically, for flow $i$, $a_{i}(n_i) (>0)$ represents the arrival time of its packet $n_i$, $l_{i}(n_i)$ its length (in bits), $A_i^P(t)$ its cumulative traffic (in packets) and $A_i(t)$ its cumulative traffic (in bits) by time $t$. \\

{\bf Remark:} An alternative way to represent a traffic process, e.g., $A(t)$ or $A^{*}(t)$, is to count the traffic at time $t$. In other words, $A(t)$ and $A^{*}(t)$ are defined up to time $t$ with $t$ included, i.e., in $[0, t]$. Since it is assumed that the system is initially empty, i.e., $A(0)=0$ and $A^{*}(0)=0$, we can change their definitions so that $A(t)$ and $A^{*}(t)$ are defined in $(0, t]$. This way of representing has also been investigated and discussed in the network calculus literature \cite{NetCal}\cite{DNC}. It can be verified that the results in this paper are not affected.

\subsection{A Calculus for Token-Bucket Traffic Characterization} \label{sec-2-2}

In network calculus, a flow is said to have an arrival curve $\alpha$ if its traffic satisfies \cite{NetCal}\cite{SNC}\cite{DNC}
$$
A(s, t) \le \alpha(t-s), \quad \forall 0 \le  s \le t.
$$
A system is said to provide a service curve $\beta$ to its input $A(t)$ \cite{NetCal}\cite{SNC}\cite{DNC}, if there holds, $\forall t \ge 0$, 
$$
A^{*}(t) \ge \inf_{0 \le s \le t}\{A(s) + \beta(t-s)\}.
$$

A special type of arrival curve has this form:
$$\alpha(\tau) = \rho \cdot \tau + \sigma$$
which was first studied by Cruz in his seminar work \cite{Cruz91ab}, and is referred to as the token-bucket $(\sigma, \rho)$ characterization of traffic:  

\begin{definition}\label{d_tb} 
The traffic of a flow is token-bucket $(\sigma, \rho)$-constrained,  
if for all $0 \le s \le t$, there holds 
\begin{equation}
A(s,t) \le \rho \cdot (t-s) + \sigma.
\end{equation}
\end{definition}

In the following, a review of the results that provide answers to (Q1) -- (Q3) for the token-bucket $(\sigma, \rho)$ traffic characterization is provided, which can be found from the network calculus literature. 

\begin{proposition}
    {\bf (Output Characterization)} If the traffic of a flow at the input is token-bucket $(\sigma, \rho)$-constrained and the system guarantees a delay bound $\Delta$ to the flow, the traffic of the flow at the output is token-bucket $(\sigma + \rho \Delta, \rho)$-constrained \cite{Cruz91ab}. 
\end{proposition}

\begin{proposition}
    {\bf (Superposition Property)} If the traffic of each individual flow $i$ in an aggregate is token-bucket $(\sigma_i, \rho_i)$-constrained, the traffic of the aggregate flow is token-bucket $(\sum_i \sigma_i, \sum_i \rho_i)$-constrained \cite{Cruz91ab}. 
\end{proposition}

\begin{proposition}\label{pr-tb-bdb}
    {\bf (Backlog and Delay Bounds)} Suppose the traffic flow to a system is token-bucket $(\sigma, \rho)$-constrained and the system provides a service curve $\beta$ to the flow. Let $\alpha(t) = \sigma + \rho \cdot t$. \\ 
    (i) The backlog at any time $t (\ge 0)$ is upper-bounded by $v(\alpha(t), \beta(t)) \equiv \sup_{t\ge 0} \{\alpha(t)- \beta(t)\}$ \cite{NetCal}. \\
    (ii) The virtual delay at any time $t (\ge 0)$ is upper-bounded by $h(\alpha(t), \beta(t))\equiv \sup_{t\ge 0}  \inf\{d: \alpha(t) \le \beta(t+d) \}$ \cite{NetCal}. \\ 
    (iii) The delay of any packet $n$ is upper-bounded by $v(\beta^{\uparrow}(t), \alpha^{\downarrow}(t+l(n))) = \sup_{t \ge 0}\{\beta^{\uparrow}(t) - \alpha^{\downarrow}(t+l(n))\}$ \cite{Jiang26}, where $\beta^{\uparrow}$ denotes the upper pseudo-inverse function of $\beta$ and $\alpha^{\downarrow}$ the lower pseudo-inverse function of $\alpha$ \cite{Chang00}\cite{NetCal}\cite{DNC}. 
\end{proposition}

\section{A Calculus for TSN Traffic Characterization}  \label{sec-3}

\subsection{TSN Traffic Characterization } 

The TSN traffic characterization is defined as  \cite{8021Q}\cite{DetNet}:
\begin{definition}\label{def-tspec}
A flow is said to conform to the TSN traffic specification $(T, S, N)$, with interval parameter $T (>0)$, maximum packet size $S(>0)$ and maximum packet number parameter $N (\ge 1)$, if during any observation interval of length $T$, the maximum packet length is not more than $S$, and the maximum number of packets from the flow in this interval is not more than $N$. 
\end{definition}

It is worth highlighting that an interval may be closed or half-open. In \cite{8021Q}\cite{DetNet}, this is not explicitly defined. In the performance analysis literature, both closed, e.g. in \cite{Cruz91ab}, and half-open intervals, e.g. in \cite{Chang00}\cite{NetCal}\cite{DNC}, have been used. In the latter case, both half-open on the left and half-open on the right are seen. {\bf If not specified, it is adopted in this paper that the interval in the TSN traffic characterization is half-open on the right} to be consistent with the definition of the traffic process $A(t)$. 

In the following, we establish a connection between the TSN model and the token-bucket model. 
Note that in the initial token-bucket model, the token rate and the bucket size use bits-per-second (bps) and bit as their units, respectively. If these units are changed to packets-per-second and packet respectively, we have a packet-version token-bucket model with parameters $(\sigma^P, \rho^P)$.    
Proposition \ref{pr-tsn2tb} relates the TSN traffic characterization to the packet-version token-bucket model and vice versa. 

\begin{proposition}\label{pr-tsn2tb}
    (i) If a flow with no concurrent arrivals has $(T, S, N)$-TSpec, it is packet-version token-bucket $(\sigma^P, \rho^P)$-constrained with $\sigma^P = N$ and $\rho^P = N/T$, i.e., $\forall 0 \le s \le t$
    \begin{eqnarray}
        A^P(s,t) &\le& \rho^P \cdot (t-s) + \sigma^P.
    \end{eqnarray}
    (ii) If a flow has a maximum packet size $l_{max}$ and its traffic is constrained by packet-version token-bucket $(\sigma^P, \rho^P)$, it conforms to $(T,S,N)$-TSpec with $T=1/\rho^P$, $S=l_{max}$ and $N=\sigma^P$.
\end{proposition}

\begin{proof}
For the first part (i), consider any period $[s, t)$. We have $A(s, t) \le N \cdot \left\lceil \frac{t-s}{T} \right\rceil \le N \cdot (\frac{t-s}{T} + 1) = \frac{N}{T}(t-s) + N$, which completes the proof.

For the second part (ii), consider any packets $m$ and $m+N$. Then, in period $[a(m), a(m+N))$, the traffic amount (in packets) is $(\sigma^P, \rho^P)$-constrained, i.e. $A^P(a(m), a(m+N)) \le \rho^P \cdot (a(m+N) - a(m))+ \sigma^P$. Note also packet $m+N$ is not in $[a(m), a(m+N))$ and there is no other arrival at $a(m)$, so $A^P(a(m), a(m+N) )= N-1$, i.e. from packet $m$ till packet $m+N-1$. Hence, we have $a(m+N) - a(m) \ge \frac{N +1 - \sigma^P}{\rho^P}$. Letting $N=\sigma^P$ gives $a(m+N) - a(m) \ge \frac{1}{\rho^P}$. Then (ii) follows from a later result, Proposition \ref{co-tspec}. 
\qed
\end{proof}

As an example, for the case $N=1$ or $\sigma^P=1$, Proposition \ref{pr-tsn2tb} provides a one-to-one mapping between the two models. Specifically, $(T, S, 1)$ gives a packet-version token-bucket $(1, 1/T)$ characterization with the maximum packet length $l_{max}=S$ from (i), which by applying (ii) returns the same $(T, S, 1)$ TSN characterization. However, for more general cases, such a one-to-one mapping is not supported by Proposition \ref{pr-tsn2tb}.\\ 

Proposition \ref{pr-tsn2tb}, particularly its part (i), is important for performance analysis because it establishes a bridge between TSN TSpec and token-bucket traffic characterization. Through this connection, a large body of existing network calculus results can be directly leveraged for TSN analysis. In particular, the relationship enables the backlog and delay analyses developed later in this paper.

\subsection{Equivalent Models}

If the interval is half-open on the right, i.e. $[s, s+\tau)$, the following equivalent model has been proved \cite{LeBoudec18}, which can also be verified with Proposition~\ref{co-tspec} and Proposition \ref{pr-tspec-2}. 

\begin{proposition}\label{pr-tsn2}
    A flow has the TSN traffic specification $(T, S, N)$ with the observation interval being half-open on the right, if and only if, $\forall n\ge m \ge 1$,  
\begin{equation}\label{eq-tspec-nuk}
a(n) -a(m) \ge T \cdot  \left\lceil \frac{n-m - N+1}{N} \right\rceil 
\end{equation}
and $S \ge l(m), \forall m \ge 1$.
\end{proposition}

In Theorem \ref{th-tspec}, we introduce a new equivalent model. The key intuition behind it is that every group of $N+1$ consecutive packets must span at least an interval of duration $T$. Consequently, no interval of length $T$ can contain more than $N$ packets, which is the requirement imposed by TSN TSpec. 

\begin{theorem}\label{th-tspec}
A sufficient and necessary condition for a flow to be in conformance with the TSN traffic specification $(T, S, N)$ {\bf where the observation interval is closed} is: For every packet $m (\ge 1)$ of the flow, there hold, $S \ge l(m)$ and  
\begin{equation}\label{eq-th-tspec-hc}
a(m+N) - a(m) > T
\end{equation}
or in other words $\forall n \ge N+1$, 
\begin{equation}\label{eq-th-tspec-e}
a(n) - a(n-N) > T. 
\end{equation}
\end{theorem}

\begin{proof}
The requirement $S \ge l(m), \forall m(\ge 1)$ is trivial. The equivalence between (\ref{eq-th-tspec-hc}) and (\ref{eq-th-tspec-e}) is also easily verified by replacing $m+N$ with $n$ in the former or  representing $n$ in the latter with $n=m+N$.  

Consider any interval $[s, s+T]$, which has length $T$, where $s (\ge 0)$ denotes the start of the interval and $s+T$ the end of the interval. Without loss of generality, suppose that $m$ is the first packet from the arrival process in this interval. 

We first prove that (\ref{eq-th-tspec-hc}) is a sufficient condition. Note that the definition of packet $m$ implies $a(m) \ge s$. Then, together with (\ref{eq-th-tspec-hc}), we have
$$
a(m+N) > a(m) + T \ge s +T 
$$
which implies that the $(m+N)$th packet is not within this interval that ends at $s+T$. In other words, there are at most $N$ packets in the interval, which are packets $m, \dots, m+N-1$. This proves the sufficient condition.

We next prove that (\ref{eq-th-tspec-hc}) is a necessary condition, or, in other words, if the process conforms to $(T, S, N)$-TSpec with the observation interval being closed, (\ref{eq-th-tspec-hc}) will hold. The necessary condition is equivalently to prove, if (\ref{eq-th-tspec-hc}) does not hold, the statement that the process conforms to $(T, S, N)$-TSpec will not be true. 
Now assume that (\ref{eq-th-tspec-hc}) does not hold, which is equivalent to assuming that there exists at least one packet making (\ref{eq-th-tspec-hc}) violated. This assumption, letting $m_o$ denote such a packet, would imply $a(m_o+N) - a(m_o) \le T$ or $a(m_o+N) \le a(m_o) + T$ . Consider the interval $[a(m_o), a(m_o)+T]$ that has length $T$. Then the packets $m_o, \dots, m_o+N$ are all in this interval, and the total number of them is $N+1$. Here, packet $m_o+N$ is within the interval due to  $a(m_o+N) \le a(m_o) + T$. In other words, if (\ref{eq-th-tspec-hc}) does not hold, there would exist at least one case that makes the statement, that is the process conforms to the TSpec, false. This completes the proof for the necessary part. \qed
\end{proof}

If the observation interval is half-open, the new equivalent model needs to be adjusted as presented in Preposition \ref{co-tspec}. Intuitively, when the interval is half-open, its length is $\epsilon (\to 0)$ shorter than when the interval is closed. Applying this to Theorem \ref{th-tspec} gives Proposition~ \ref{co-tspec}, and a more formal proof can be readily derived by following the steps used in the proof of Theorem \ref{th-tspec}. As a remark, Proposition~\ref{co-tspec} holds regardless of whether the observation interval is half-open on the left or half-open on the right. 

\begin{proposition}\label{co-tspec}
           A flow conforms to the TSN traffic specification $(T, S, N)$ where {\bf the observation interval is half-open}, if and only if, $\forall m \ge 1$, $S\ge l(m)$ and  
        \begin{equation}\label{eq-co-ts1}
        a(m+N) - a(m) \ge T
        \end{equation}
        or in other words $\forall n \ge N+1$, 
        \begin{equation}\label{eq-co-ts2}
        a(n) - a(n-N) \ge T. 
\end{equation}
\end{proposition}

Comparing (\ref{eq-co-ts1}) with (\ref{eq-tspec-nuk}), the former is clearly much simpler in representation. An immediate question is: Are they equivalent? The following result provides the answer.

\begin{proposition}\label{pr-tspec-2}
For a flow in compliance with the TSN traffic specification $(T, S, N)$ {\bf where the observation interval is half-open on the right}, conditions (\ref{eq-co-ts1}) and (\ref{eq-tspec-nuk}) are equivalent, i.e., both are sufficient and necessary conditions for conforming to the $(T, S, N)$-TSpec. 
\end{proposition}

\begin{proof}
We first prove that (\ref{eq-tspec-nuk}) implies (\ref{eq-co-ts1}). For this, letting $n=m+N$ in (\ref{eq-tspec-nuk}) gives
$$
a(m+N) - a(m) \ge T \cdot \left\lceil \frac{1}{N} \right\rceil = T.
$$ 

We next prove that (\ref{eq-co-ts1}) implies (\ref{eq-tspec-nuk}). For $\forall n \ge m \ge 1$, we have from (\ref{eq-co-ts1}):
 \begin{eqnarray}
a(n) - a(m) &=& \left ( a(m + I \cdot N) - a(m) \right ) + \left ( a(n) - a(m + I \cdot N) \right )\nonumber 
 \end{eqnarray}
 where $I = \lfloor \frac{n-m}{N} \rfloor$. 
 
There are two possible cases: $I=0$ and $I \ge 1$.  The former case implies that $0 \le n-m \le N-1$. Then, on the one hand we have $a(n) - a(m) \ge 0$ and on the other hand, $T \cdot\left\lceil \frac{n-m - N+1}{N} \right\rceil =0$, so for this case, i.e. $I=0$, $a(n) - a(m) \ge T \cdot \left\lceil \frac{n-m - N+1}{N} \right\rceil $. 

For the case of  $I \ge 1$, since $I = \lfloor \frac{n-m}{N} \rfloor \le \frac{n-m}{N}$, we have $n \ge m + I \cdot N$, which implies $a(n) \ge a(m + I \cdot N)$. Together with (\ref{eq-co-ts1}), we have 
 \begin{eqnarray}
a(n) - a(m)  &\ge& a(m + I \cdot N) - a(m) \nonumber\\
&=& \sum_{i=1}^{I} 
\left[ a(m+ i \cdot N) - a(m +(i-1) \cdot N) \right]
\label{tmp-36} \nonumber\\
&\ge& I \cdot T  = \left\lfloor \frac{n-m}{N} \right\rfloor \cdot T \label{tmp-37} \nonumber\\
&=& T \cdot \left\lceil \frac{n-m-N+1}{N} \right\rceil 
 \end{eqnarray}
where the last step is from the property of ceiling and floor functions: $\lfloor \frac{N}{M} \rfloor = \lceil \frac{N-M+1}{M} \rceil$ for any positive integer $M$. This proves that (\ref{eq-co-ts1}) implies (\ref{eq-tspec-nuk}). \qed
\end{proof}

\subsection{Basic Properties}

In this subsection, several basic properties of the TSN traffic characterization are proved, which correspond to those for the token-bucket traffic characterization reviewed in Section \ref{sec-2-2} and provide answers to (Q1) -- (Q3). 

\begin{proposition}\label{pr-out}
    {\bf (Output Characterization)} Consider any FIFO system with bounded packet delay $\Delta$. If its input conforms to $(T, S, N)$-TSpec, its output has TSN traffic characterizations \\
(i) $(T - \Delta, S, N)$ if $\Delta < T$, and \\
(ii) $(T, S, N')$ where $N'=N+\left\lceil \frac{\frac{\Delta}{T} + N-1}{N}\right\rceil N$. 
\end{proposition}

\begin{proof}
Since the maximum packet size does not change, the output has the same maximum packet size $S$ as the input. 

Let $d(n)$ denote the departure time of packet $n$ from the FIFO system. For any packet $m \ge 1$, $d(m) \le a(m) + \Delta$ under the assumption that the delay of any packet is upper-bounded. In addition, the departure cannot be earlier than the arrival, so $d(m+N) \ge a(m+N)$. Together, we obtain 
$$
d(m+N) - d(m) \ge a(m+N) - [a(m) + \Delta] = a(m+N) - a(m) - \Delta \ge T - \Delta.
$$
which proves (i) based on Proposition~\ref{co-tspec}.

For (ii), there holds $d(m+N') - d(m) = a(m+N')-a(m) + (d(m+N') - a(m+N')) - (d(m)-a(m)) 
\ge a(m+N')-a(m) -\Delta$. Then, with the equivalent model in Proposition \ref{pr-tsn2} which gives $a(m+N')-a(m) \ge T \cdot \left\lceil \frac{N'-N+1}{N} \right\rceil $, we have 
\begin{eqnarray}
    d(m+N') - d(m) 
&\ge& T \cdot \left\lceil \frac{N'-N+1}{N} \right\rceil -\Delta \nonumber\\
&=& T \cdot \left\lceil \frac{\left\lceil \frac{\frac{\Delta}{T} + N-1}{N}\right\rceil N +1}{N} \right\rceil -\Delta  \nonumber\\
&\ge& T \cdot \left\lceil 1+ 
\frac{\Delta}{T} 
\right\rceil - \Delta \ge T
\end{eqnarray}
which proves the second part based on Proposition~\ref{co-tspec}. \qed
\end{proof}

As a remark, the output has two distinct TSN traffic characterizations. A comparison of them is provided and discussed in the case study in Section \ref{sec-4}.

\begin{proposition}\label{pr-sup}
     {\bf (Superposition Property)} Suppose the traffic of each individual flow $i$ in an aggregate conforms to $(T_i, S_i, N_i)$-TSpec. The traffic of the aggregate flow has TSN traffic characterizations: \\
(i) $(T_{min}, S, N)$ with $T_{min} = \min_i\{T_i\}$ and $N = \sum_i N_i$, \\
(ii) $(T_{max}, S, N_{max})$ with $T_{max} = \max_i\{T_i\}$ and $N_{max} = \sum_i \left\lceil \frac{T_{max}}{T_i}\right\rceil N_i$, and \\
(iii) $(T_{lcm}, S, N_{lcm})$ with $ T_{lcm} = LCM \{T_i\}$ and $N_{lcm} = \sum_{i} \frac{T_{lcm}}{T_i} N_i$, \\
where  $S=\max_i\{S_i\}$ and $LCM_i\{T_i\}$ denotes the least common multiple (LCM) of $T_i, \forall i$. 
\end{proposition}
 
 \begin{proof}
Since $S_i$ denotes the maximum packet size of flow $i$, $S=\max_i\{S_i\}$ is hence the maximum packet size of the aggregate. 
 
For (i), consider any period of length $T_{min}$. Since it is not greater than $T_i, \forall i$, then for each flow $i$, there are at most $N_i$ packets of the flow in this period. Together, the aggregate has at most $N_{min}=\sum_i N_i$ packets in the period, which proves (i). The same intuition can be applied to verify (ii) and (iii). 
In the remaining, a more formal proof is provided for (i), and the same technique can be extended for (ii) and (iii). 

We prove (i) by induction. Let $I$ denote the number of individual flows in the aggregate. Consider the base case  $I=2$, and $a(m+N) - a(m)$, $\forall m \ge 1$ for the aggregate. Let $N=N_1 + N_2$. Without loss of generality, suppose packet $m$ belongs to flow $1$ and is its $m_1$-th packet. Then, from $a(m)$ to $a(m+N)$, because $N=N_1 + N_2$, there are two cases covering all possible combinations of packets from the two flows: 

Case 1: There are at least $N_1$ packets from flow $1$, implying at most $N_2$ packets from flow $2$.  

Case  2: There are at least $N_2$ packets from flow $2$.
 
Under Case 1, since packet $m$ of the aggregate is the packet $m_1$ of flow 1 and $N \ge N_1$ we must have 
$$
a(m+N)  \ge a_1(m_1+N_1) 
$$ 
and hence
\begin{eqnarray}
a(m+N) - a(m) & \ge & a_1(m_1+N_1)  - a(m) \nonumber\\
&=&  a_1(m_1+N_1)  - a(m_1) \ge T_1 \ge T_{min}
 \end{eqnarray}

Under Case 2, let $m_2$ denote the first packet from flow $2$ in the period of $a(m)$ to $a(m+K)$. Then, we must have  
$$
a_2(m_2)  \ge a(m).
$$
In addition, the condition of Case 2 also implies 
$$
a_2(m_2+N_2)  \le a(m+N).
$$
Hence, the following is obtained
\begin{eqnarray}
a(m+N) - a(m) & \ge & a_2(m_2+N_2)  - a_2(m_2) \ge T_2 \ge T_{min}. 
 \end{eqnarray}
 
Combining both Case 1 and Case 2, with the equivalent definition of the TSN traffic characterization shown in Proposition~\ref{co-tspec}, the base case is proved. For the induction step, consider $I+1$ flows. By treating the first $I$ flows as one flow and applying the same proof to the $I$-aggregate flow and flow $I+1$, the induction step is verified. This completes the proof of (i). \qed 
 \end{proof}
 
{\bf Remark:} In contrast to the token-bucket model whose superposition property is represented by a single expression, for TSN TSpec, we have multiple ways to characterize the traffic of the aggregate shown in Proposition \ref{pr-sup}. In the case study conducted in Section \ref{sec-4}, a brief discussion of this is provided. 
It is worth noting that the aggregate characterizations in Proposition \ref{pr-sup} prioritize closure and analytical tractability. In general, they are not necessarily tight. Deriving tighter aggregate TSN characterizations remains an open problem for future research.\\

Finally, by applying the token-bucket characterization of the TSN model to the backlog and delay bounds for token-bucket-constrained traffic shown in Proposition \ref{pr-tb-bdb}, we also obtain backlog and delay bounds for a flow conforming to the TSN TSpec. The results are summarized in Proposition \ref{pr-bdb}.

\begin{proposition}\label{pr-bdb}
     {\bf (Backlog and Delay Bounds)} Suppose the traffic flow to a system conforms to the $(T, S, N)$-TSpec and the system provides a service curve $\beta$ to the flow. Let $\alpha(t) = \frac{N \cdot S}{T} t + N \cdot S$. \\ 
    (i) The backlog at any time $t$ is upper-bounded by $v(\alpha(t), \beta(t))$. \\
    (ii) The virtual delay at any time $t$ is upper-bounded by $h(\alpha(t), \beta(t))$. \\ 
    (iii) The delay of any packet $n$ is upper-bounded by $v(\beta^{\uparrow}(u), \alpha^{\downarrow}(u+S))$. 
\end{proposition}

\section{Case Study and Discussion} \label{sec-4}

To demonstrate the applicability of the developed TSN calculus and gain insight into its practical implications, we conduct a case study based on a TSN configuration example from IEEE 802.1Q Annex U \cite{8021Q}. The case study illustrates the superposition property, backlog and delay bound analysis, and output characterization results derived in the previous section.

\subsection{The Configuration}
The configuration is as described in Annex U: TSN Configuration Examples \cite{8021Q}. Specifically, a TSN time-aware Talker transmits packets on a link. 
The transmission rate of the link is $c=100$ Mb/s. The Talker has two queues, a TSN queue and a non-TSN queue, where the TSN queue has non-preemptive strict priority over the non-TSN queue. When the TSN queue is enabled, a non-TSN frame can interfere and this interference is up to $\tau_{interference} = 123.36$ $\mu s$, assuming maximum frame size 1522 bytes for the non-TSN queue. 

The Talker supports two TSN streams, Stream J and Stream K. Both generate frames periodically, i.e. $N_J=1$ and $N_K=1$. For Stream J, the frame size is $S_J = 1500$ bytes and the interval is $T_J=500$ $\mu s$. For Stream K, the frame size is $S_K = 1000$ bytes, and the interval is also $T_K=500$ $\mu s$. In summary, Stream J has TSN traffic characterization $(T_J, S_J, N_J)$, and Stream K has TSN traffic characterization $(T_K, S_K, N_K)$, where $T_J=T_K = 500$ $\mu s$, $S_J=1500$ bytes, $S_K=1000$ bytes, and $N_J=N_K=1$. 

The two streams form the input to the TSN queue, and hence for the TSN queue, $l_{max} = 1500$ bytes and $l_{min} = 1000$ bytes. 

\subsection{Basic properties}

\subsubsection{Superposition}

In this case, both flows have the same interval characterization, i.e. $T_J=T_K=500 \mu s$. It is easily verified that for their aggregate, $T_{min}=T_{max}=T_{lcm}= 500 \mu s$, $S=\max\{S_J, S_K\}= 1500$ bytes, $N=N_{max}=N_{lcm}=2$. So, the aggregate conforms to (500 $\mu s$, 1500 bytes, 2)-TSpec. 

\subsubsection{Backlog and delay bounds}
It can be verified that Stream J is token-bucket $(\sigma_J, \rho_J)$-constrained with $\rho_J= 24$ Mb/s and $\sigma_J=S_J=1500$ bytes, and Stream K is token-bucket $(\sigma_K, \rho_K)$-constrained with $\rho_K= 16$ Mb/s and $\sigma_K=S_K = 1000$ bytes. The aggregate input to the TSN queue is token-bucket $(\sigma, \rho)$-constrained with $\rho = \rho_J + \rho_K = 40$ Mb/s and $\sigma=S_J+S_K=2500$ bytes. 

Since the TSN queue is at the highest priority level, the system provides to the TSN-queue rate-latency service curve $\beta(t) = R(t-T)^{+}$ with $R=c= 100$ Mb/s and $T=\frac{l_{max}}{c} + \tau_{interference} = 243.36$ $\mu s$ (e.g. see Sec. 2.4.2, \cite{Chang00}). With these, the backlog, virtual delay and packet delay are all bounded for the aggregate, Stream J and Stream K. The bounds are summarized and compared in Table \ref{tb-1}. For simplicity, the backlog bound for the aggregate is used for both Stream J and Stream K. For delay, it is known that under FIFO, all flows have similar queueing delay performance. This is confirmed in Table \ref{tb-1}. The improvement in packet-delay bound for Stream J is due to its smaller packet size and hence shorter transmission time, and the packet-delay bound is end-to-end, i.e. on the sum of queueing delay and transmission time. 

\begin{table}[th!]
\caption{Backlog and delay bounds for the aggregate, Stream J and Stream K} \label{tb-1}
\centering
\begin{tabular}{|l|c|c|c|}
\hline
Flow & Backlog bound (bytes)  & Virtual-delay bound ($\mu s$)  &  Packet-delay bound ($\mu s$)\\ \hline \hline
Aggregate & 4022 & 443.36   & 363.36 \\ \hline 
Stream J & 4022 & 443.36 & 323.36 \\ \hline
Stream K & 4022 & 443.36 & 363.36 \\ \hline 
\end{tabular}%
\nop{
\begin{tabular}{|l|c|c||c|c|c|}
\hline
Stream & $\Delta^{New}$ &  $\Delta$ &  $\Delta_i^{NEW}$ & $\Delta_i^{MSL19}$ & $\Delta_i^{MSL23}$\\ 
\hline\hline  
 J & 363.36 & 443.36   & 323.36 & 443.36  & 443.36  \\ \hline 
 K & 363.36 & 443.36   & 363.36 & 443.36  & 443.36 \\ \hline 
\end{tabular}%
}
\end{table}

\subsubsection{Output Characterization}
The output TSN characterization results are summarized in Table \ref{tb-2}. They are derived by applying Proposition \ref{pr-out} and the packet delay bounds in Table \ref{tb-1}. Note that, for each flow, there are two different TSN characterizations.  

\begin{table}[th!]
\caption{Output characterization for the aggregate, Stream J and Stream K} \label{tb-2}
\centering
\begin{tabular}{|l|c|c|c|}
\hline
Flow & Interval time ($\mu s$) & Max. size (bytes) &  Max. pkt number\\ \hline \hline
Aggregate & 136.64 & 1500   & 2\\ \hline 
Stream J & 176.64 & 1500 & 1 \\ \hline
Stream K & 136.64 & 1000 & 1 \\ \hline \hline
Aggregate & 500 & 1500   & 4\\ \hline 
Stream J & 500 & 1500 & 2 \\ \hline
Stream K & 500 & 1000 & 2 \\ \hline 
\end{tabular}%
\end{table}

\subsection{Discussion}

In Proposition \ref{pr-sup}, three TSN characterizations are introduced for the aggregate flow, depending on if the interval time parameter has the same value for all constituent flows. For the considered case, due to that both flows have the same interval time value, the three characterizations become the same. It is worth highlighting that if the individual flows have different interval values, the three TSN characterizations from Proposition \ref{pr-sup} will be different.

In backlog and delay bound analysis, we have first applied token-bucket characterization to each individual flow, then adopted token-bucket characterization for the aggregate flow, and finally applied network calculus backlog and delay bound results. As a comparison, for the aggregate, there is an alternative approach. It is to first apply the superposition property to obtain the TSN characterization of the aggregate, i.e., (500 $\mu s$, 1500 bytes, 2). Then we can find its token-bucket characterization based on this aggregate TSN characterization, which gives $\sigma' =3000$ bytes and $\rho' = 48$ Mb/s. Furthermore, we can apply network calculus backlog and delay bound results which give 4522 bytes for backlog, 483.36 $\mu$s for virtual delay and 403.36 $\mu$s for packet delay. Comparing with the counterparts in Table \ref{tb-1}, the bounds from the alternative approach are looser.  

For the output characterization, at a first glance, it seems Table \ref{tb-2} only indicates that the output can have two different TSN characterizations. Then either of them may be applied in the next hop of a flow. Interestingly, they have a much severe implication for possibly being applied in the next hop. Take Stream J as an example. Suppose the next hop is given its output characterization from the Talker. Then, to this hop, the rate of Stream J would become 68 Mb/s if the characterization keeps the maximum packet number or 48 Mb/s if the characterization keeps the interval time! Both rate values are significantly higher than its actual data rate that is 24 Mb/s. 
More specifically, the inferred rate increases from the actual value of 24 Mb/s to either 48 Mb/s or 68 Mb/s, corresponding to inflation factors of 2.0 and 2.83, respectively. Similar inflation occurs for Stream K and the aggregate flow. This observation highlights that repeated application of output characterization can rapidly lead to overly conservative traffic descriptions. 

Two most important observations from the case study are as follows. One is that, for backlog and delay bound analysis, it is preferable to apply token-bucket characterization earlier in the process. 
Another is that TSN TSpec parameters may progressively deteriorate under forwarding. While the traffic remains characterizable by TSN TSpec, the resulting parameters can become significantly more conservative than the actual traffic characteristics, potentially leading to substantial overestimation of required resources. An implication is that re-shaping the output to the initial TSN characterization is not just another traffic control approach but may be necessary. 

As a remark, the TSN traffic specification may optionally be extended to include additional parameters, such as {\em a minimum packet size} and {\em a minimum number of packets in one interval} \cite{rfc9016}. They may be integrated in the basic properties derived for the TSN traffic characterization. For example, the minimum packet size remains unchanged in the output for Proposition \ref{pr-out}. 

\section{Conclusion}  \label{sec-5}

This paper investigated fundamental traffic characterization problems for traffic specification (TSpec) in Time-Sensitive Networking (TSN) and Deterministic Networking (DetNet). We first derived a new equivalent characterization of TSN TSpec, which simplifies conformance verification and analytical development. Based on this result, we established an output characterization property and showed that the output of a TSN-constrained flow traversing a bounded-delay server can be characterized by a TSN TSpec with updated parameters. We further studied the superposition of TSN-constrained flows and derived aggregate traffic characterizations for multiplexed traffic streams. The obtained results provide a theoretical foundation for analyzing TSN and DetNet networks with TSN-native traffic models.

The case study further revealed an important practical implication. Although TSN TSpecs are preserved under forwarding and aggregation, the resulting parameters may become increasingly conservative as traffic traverses the network. This conservatism can inflate estimated resource requirements and adversely affect admission decisions at downstream nodes. The results therefore suggest that traffic reshaping is not merely an optional traffic-control mechanism but may be necessary to maintain efficient resource utilization in TSN and DetNet networks.

\section*{Acknowledgement}
 The author would like to thank Jean-Yves Le Boudec for a discussion on paper \cite{Jiang18}, which triggered the formulation of the equivalent model in Theorem \ref{th-tspec}.   

\bibliographystyle{unsrt}
\bibliography{Valuetools26} 
\end{document}